\documentclass[prr,twocolumn,showpacs,aps,superscriptaddress, longbibliography]{revtex4-1}
 
\usepackage{amsmath}
\usepackage{amssymb}
\usepackage{amsmath}
\usepackage{txfonts}
\usepackage{graphicx}
\usepackage{epsfig}
\usepackage[T1]{fontenc}
\usepackage{times}
\usepackage{color} 
\usepackage{xspace}
\usepackage{amssymb,amsmath}
\usepackage{amsbsy}
\usepackage{braket}
\usepackage{tabularx}
\usepackage{mathtools}          

\usepackage{ifpdf}
\ifpdf
\usepackage{epstopdf}
\fi

\usepackage[unicode]{hyperref}
\hypersetup{
	pdftitle={spinnorBEC},
	pdfauthor={Yu},
	pdfsubject={magneticdominawall},
	colorlinks=true,
	linkcolor=blue,
	citecolor=blue,
	filecolor=black,
	urlcolor=blue
}
\def\urlprefix{}
\def\url#1{}

\usepackage{bm}
\usepackage{color}

\newcommand{\be}{\begin{equation}}
	
\newcommand{\ee}{\end{equation}}

\newcommand{\bea}{\begin{eqnarray}}

\newcommand{\eea}{\end{eqnarray}}

\newcommand{\nn}{\nonumber }

\newcommand{\abs}[1]{{|{#1}|}}

\begin{document}

\title{Hydrodynamics of Quantum Vortices on a Closed Surface}

\author{Yanqi Xiong}
\affiliation{Graduate School of China Academy of Engineering Physics, Beijing 100193, China}
\author{Xiaoquan Yu}
\email{xqyu@gscaep.ac.cn}

\affiliation{Graduate School of China Academy of Engineering Physics, Beijing 100193, China}
\affiliation{Department of Physics, Centre for Quantum Science, and Dodd-Walls Centre for Photonic and Quantum Technologies, University of Otago, Dunedin, New Zealand}

\begin{abstract}
We develop a neutral vortex fluid theory on closed surfaces with zero genus. The theory describes collective dynamics of many well-separated quantum vortices in a superfluid confined on a closed surface. Comparing to the case on a plane, the covariant vortex fluid equation on a curved surface contains an additional term proportional to Gaussian curvature multiplying the circulation quantum. This term describes the coupling between topological defects and curvature in the macroscopic level. For a sphere, the simplest nontrivial stationary vortex flow is obtained analytically and this flow is analogous to the celebrated zonal Rossby-Haurwitz wave in classical fluids on a nonrotating sphere. For this flow the difference between the coarse-grained vortex velocity field and the fluid velocity field generated by vortices is solely driven by curvature and vanishes in the corresponding vortex flow on a plane when the radius of the sphere goes to infinity.

\end{abstract}

\maketitle

\section{Introduction}
Fluids on curved surfaces exhibit rich phenomena which are absent on a plane. The interplay between  geometry, topology and fluid dynamics has been explored  extensively in diverse platforms, including  quantum Hall liquids~\cite{son2013newtoncartan, Wiegmann2013, Cho2014, Gromov14, Gromov2015prl, CAN2015},  active matter~\cite{Mickelin18, Shankar2017, Rank2021},  classical fluids~\cite{bogomolov1977dynamics, Reuther2015, Reuther2018, SAMAVAKI2020}, and soft materials~\cite{C6SM00194G, PhysRevE.69.021503, ROWER2022110994}.

The coupling between geometric potentials induced by curvature and quantum vortices plays an essential role in determining  properties of superfluids on a curved surface~\cite{Vitelli2004, Turner2010}.  For a superfluid film,  a curved surface is realized by the underlying  substrate~\cite{Turner2010}.   Recent experimental advances in Bose-Einstein condensates (BECs) in  International Space Station~\cite{aveline2020observation}  now allow ultracold atomic bubbles~\cite{carollo2022observation},  providing a promising  possibility to investigate  a bubble trapped superfluid experimentally.   Motivated by the experimental progress,  research interests on few body vortex dynamics on curved surfaces have been renewed~\cite{PhysRevA.103.053306, PhysRevA.105.023307, PhysRevA.102.043305}, adding different perspectives on a more mathematical treatment of point vortex dynamics on curved surfaces~\cite{Hally, Kimura99, dritschel2015motion}.  However, the effects of curvature and topology on collective dynamics of quantum vortices  remain unexplored, motivating us to consider  vortex fluids on curved surfaces.  Furthermore,  static vortex distributions influenced by curvature remains a challenge~\cite{Turner2010}, especially when the vortex number is large.  Examining stationary solutions of such vortex fluid equations would provide a feasible  way to tackle  this problem.

A vortex fluid is a coarse-grained model for a system consisting of a large number of point vortices and its dynamical equations  describe  collective dynamics of well-separated  quantum vortices at  large scales~\cite{PhysRevLett.113.034501,Yu:2017siv}. The theory reveals several emergent  properties. For instance,  a binary vortex fluid is compressible~\cite{Yu:2017siv} while a chiral vortex fluid is incompressible~\cite{PhysRevLett.113.034501}; there exists an odd  viscous tensor and  the circulation quantum plays the role of the  nondissipative odd  viscosity coefficient. 
The  theory also  predicts a universal  long-time dynamics of the vorticity distribution in a dissipative superfluid and this prediction  has been verified in experiments~\cite{Oliver2020}.  
However, on a finite region with boundaries,  boundary conditions are difficult to incorporate in general,  hence a closed surface is a better venue for vortex fluids.  Vortex fluids are also closely related to  quantum Hall liquids~\cite{VortexfluidHalleffect} and  fractons~\cite{doshi21,Grosvenor2021}.   

In this paper we develop a vortex fluid theory on orientable closed surfaces with zero genus.  
For a closed surface,  the total voticity must vanish and  hence we  consider binary vortex fluids containing equal number of vortices and anti-vortices.   On a plane,  the momentum flux tensor of the vortex fluid contains an emergent odd viscous tensor and a quantum pressure like stress tensor~\cite{Yu:2017siv},  preventing applying the minimal coupling principle directly to derive the covariant vortex fluid equation  on a curved surface.  We overcome this difficulty by introducing an auxiliary tensor  which is mathematically equivalent to the original  momentum flux tensor, however, is more readily amenable for applying the minimal coupling substitution.  After the minimal coupling substitution and rewriting the equation in terms of  the original momentum flux tensor, we obtain the vortex fluid equation on a closed surface in isothermal coordinates.   The emergent curvature term plays the role of  a source  term in the  vortex fluid equation and hence might be referred to as {\em curvature anomaly}. The generalized relation between the superfluid velocity field generated by the vortices and  the coarse-grained vortex velocity field  induces the equation of motion (EOM) of point vortices on closed surfaces, verifying  the minimal coupling approach.  A connection between the odd viscous tensor and Euler characteristic of the closed surface is obtained.
For a  sphere,   an exact stationary vortex flow solution determined  by Gaussian curvature is found, whose vorticity exhibits the profile of a vortex-dipole in spherical coordinates and its velocity distribution has the profile of a  Kaufmann vortex in stereographic coordinates.
It should be noted that the obtained vortex fluid equation holds also for infinitely large curved surfaces, where the vortex system does not have to be neutral.

\section{Quantum vortices and vortex fluids on a plane}
In a superfluid,  the circulation of a vortex is quantized  in units of circulation quantum $\kappa \equiv2 \pi \hbar/m$~\cite{BECbook}, and the vorticity has a singularity at the vortex core $\mathbf{r}_i$: $\omega(\mathbf{r})=\nabla \times \mathbf{u}=\kappa \sigma_i \delta(\mathbf{r}-\mathbf{r}_i)$ with sign $\sigma_i=\pm 1$ for singly charged vortices.  Here $m$ is the atomic mass and $\mathbf{u}$ is the fluid velocity generated by the vortex at $\mathbf{r}=\mathbf{r}_i$.  This quantization arises from the single-valuedness  of the macroscopic  superfluid wave function.  It ensures that the vorticity of a quantum vortex concentrates around the core region in the dynamics, which is not the general case for classical fluids~\cite{EyinkRMP}.  Hence when the mean separation between quantum vortices is much larger than the vortex  core size $\ell$, the point vortex model  governs the  dynamics of quantum vortices~\cite{Onsager1949,EyinkRMP,novikov1975,Aref1999},  provided vortex annihilation can be neglected.  In this regime,  a superfluid  at low temperatures is  nearly  incompressible. 

Let us introduce complex coordinates $z=x^1+i x^2 $, $\partial_z \equiv(\partial_1-i\partial_2)/2,$ $\partial_{\bar{z}}\equiv(\partial_1+i\partial_2)/2$ and  complex velocity $u \equiv u^1-iu^2$.  For a  system containing $N_+$ singly-charged quantum vortices and $N_-$ anti-vortices, 
the superfluid velocity $u$ generated by these  vortices and the vortex velocity $v_i\equiv d \bar{z}_i/dt$ read 
\bea
\label{Kirchhoff}
u=-\frac{1}{2\pi}\sum^N_{j=1} \frac{i \kappa \sigma_j}{z-z_j}, \quad 
v_i=-\frac{1}{2\pi}\sum^N_{j,j\neq i} \frac{i \kappa \sigma_j}{ z_i(t)-z_j(t)},
\eea
where $u=2i \partial_z \psi$,  the stream function 
$\psi(z)=-\kappa /2\pi\sum_i \sigma_i \log\left|(z-z_i)/\ell\right|$, and $N=N_++N_-$ is the total number of vortices. 
The vorticity is $\omega(\mathbf{r})=\kappa \sum_i \sigma_i \delta(\mathbf{r}-\mathbf{r}_i)$.
The above fluid velocity $u$ appears to be  a singular solution of incompressible two-dimensional (2D) Euler or Hemlholtz equation~\cite{Kirchhoff1876} : $\partial_t \omega +\mathbf{u}\cdot\nabla \omega=0$,  which describes 2D nonviscous incompressible classical fluids .

In the point vortex regime,   the slow motion of  vortices is nearly  decoupled from fast degree of freedom--acoustic  modes.  In this regime, a  large number of  well-separated  quantum vortices are  almost   isolated and can be treated  as a fluid~\cite{Wiegmann, Yu:2017siv}.  On a plane, the corresponding  hydrodynamical equation is~\cite{Yu:2017siv}
\bea
&&\partial_t(\rho v^\alpha)+\partial_\beta T^{\alpha\beta}
+\rho\partial^\alpha p=0,
\label{vortexfluid}
\eea
where  $\rho(\mathbf{r})\equiv\sum_i\delta(\mathbf{r}-\mathbf{r}_i)$ is vortex number density, $\sigma(\mathbf{r})\equiv\sum_i\sigma_i \delta(\mathbf{r}-\mathbf{r}_i)=\kappa^{-1}\omega$ is  vortex charge density,  $v^{\alpha}$ is vortex velocity field defined as $\rho v^{\alpha} \equiv \sum_{i} \delta (\mathbf{r}-\mathbf{r}_i) v^{\alpha}_i$,  $p$ is the fluid pressure, 
$T^{\alpha \beta}=\rho v^\alpha v^\beta - \Pi^{\alpha\beta}$ is the momentum flux tensor,  and
\bea
\Pi^{\alpha \beta}=-\eta^2\sigma\partial^\beta\left(\frac{1}{\rho}\partial^\alpha\sigma\right)-8\pi\eta^2\sigma^2\delta^{\alpha\beta}-\sigma\tau^{\alpha\beta}
\label{momentumflux}
\eea
is the emergent Cauchy stress tensor  with $\eta = \kappa/8\pi$.  
This emergent tensor $\Pi^{\alpha \beta}$ is absent in the Euler equation and describes emergent macroscopic effects of discrete quantum vortices.
In particular, 
\bea
\label{tau}
\tau^{\alpha \beta}=-\eta \left(\epsilon^\alpha_\gamma \partial^\beta v^\gamma+\epsilon^\beta_\gamma \partial^\gamma v^\alpha\right) 
\eea
is the nondissipative  odd viscous tensor and  $\eta$ is identified as  the  odd viscosity coefficient.  Here $\epsilon^1_2=1, \epsilon^2_1=-1$, and $\epsilon^1_1=\epsilon^2_2=0$.  
The presence of $\tau^{\alpha \beta}$  in  Eq.~\eqref{momentumflux} is due to that in a vortex system  the parity symmetry is broken,  namely $\eta \rightarrow -\eta$ under the parity transformation $(x^1,x^2) \rightarrow  (-x^1,x^2) \, \text{or}  \, (x^1,-x^2)$.   The odd viscosity effects in 2D fluids are very rich ~\cite{avron1998odd, oddviscosity}  and  have been investigated in  quantum Hall systems~\cite{viscosityQHF, Hoyos2012, Bradlyn2012, Abanov2013}, chiral active matter~\cite{Souslov2019,banerjee2017odd,Xin2022},
chiral superfluids~\cite{Hoyos2014}, 2D vortex matter~\cite{Wiegmann, Yu:2017siv, Bogatskiy2019,Bogatskiy_2019, Sergej2018} and classical fluids~\cite{Ganeshan2017,Monteiro2021}.  

The vortex fluid  theory describes emergent collective dynamics of well-separated quantum votices at large scales and is valid whenever the point vortex model (PVM) Eq.~\eqref{Kirchhoff} is applicable.  
Vortex annihilations due to collisions, dissipative vortex dipole decay and boundary loss would modify the PVM.  For a low temperature BEC containing many vortices, annihilations due to collisions limit dominantly the PVM approach.  
However,  the collision rate depends on the vortex distribution~\cite{Yu:2017siv} and for largely vorticity-polarized states, including the vortex shear flow~\cite{Yu:2017siv}, Onsager clustered states~\cite{Onsager1949, Matt2014, clusteringYu, science.aat5718, DecayingQTBillam, AnghelutaPRE}, and the enstrophy cascade~\cite{Matt2017},   vortex number losses are negligible.
\section{Vortex fluids  on closed  surfaces} 
The  wisdom on deriving
laws of physics in curved spacetime  from those in flat spacetime is the so-called minimal coupling (MC) principle. For our situation,  it means the following substitution: 
\bea
\delta_{\mu \nu} \rightarrow g_{\mu \nu} ; \quad \partial_{\mu}  \rightarrow \nabla_{\mu}  ,
\label{minimalcoupling}
\eea
where $g_{\mu \nu}$ the metric on the surface, and $\nabla_{\mu}$  is Levi-Civita covariant derivative.  When acting a vector field $V^{\nu}$,  
\bea
\nabla_\mu V^\nu=\partial_\mu V^\nu+\Gamma^\nu_{\mu\lambda}V^\lambda,
\eea
where
 \bea
 \Gamma^\nu_{\mu\lambda}=\frac{1}{2}g^{\nu \rho}\left (\frac{\partial g_{\rho \mu}}{\partial x^{\lambda}}+\frac{\partial g_{\rho \lambda}}{\partial x^{\mu}}-\frac{\partial g_{ \mu\lambda}}{\partial x^{\rho}}\right)
 \eea
  is the connection coefficient--Christoffel symbol.  The second covariant derivatives do not commute,  namely 
\bea
(\nabla_\alpha \, \nabla_\beta-\nabla_\beta \, \nabla_\alpha ) V^\mu=R^\mu_{\nu \alpha \beta} V^\nu,
\eea
 where  $R^\mu_{\nu \alpha \beta}$ is Riemann curvature tensor.

Unless specified,  in the  following we use isothermal coordinates 
\bea
ds^2=g_{\mu \nu}dx^{\mu}dx^{\nu}=h(x^1,x^2)[(dx^{1})^2+(dx^{2})^2],
\eea
namely, $g_{12}=g_{21}=0$ and $g_{11}=g_{22}=h(x^1,x^2)$, where $h(x^1,x^2)$ is a positive function and  exists locally for 2D surfaces~\cite{chern1955elementary}.  In  isothermal coordinates, calculations are considerably  simplified. For instance,  $g^{\alpha \beta}= \delta^{\alpha \beta} h^{-1}$ and  $v^{\alpha}=g^{\alpha\beta} v_{\beta}=h^{-1} v_{\alpha}$.

We define the vortex number density and vortex charge density on a curved surface as
\bea
\rho(x^{\mu})=\frac{1}{\sqrt{\det g_{\mu \nu}}}\sum_i\delta(x^{\mu}-x^{\mu}_i),\\
\sigma(x^{\mu})=\frac{1}{\sqrt{\det g_{\mu \nu}}}\sum_i\sigma_i\delta(x^{\mu}-x^{\mu}_i).
\eea
The assumption of absence of  vortex annihilation ensures the the following continuity equations: 
\bea
\partial_t\rho+\nabla_{\mu} J_n^{\mu}=0, \quad \partial_t\sigma+\nabla_{\mu} J_c^{\mu}=0
\eea
where 
\bea
J^{\mu}_n&&=\frac{1}{\sqrt{\det g_{\mu \nu}}}\sum_i\delta(\mathbf{r}-\mathbf{r}_i)v^{\mu}_i\equiv\rho v^{\mu},\\
J^{\mu}_c&&=\frac{1}{\sqrt{\det g_{\mu\nu}}}\sum_i\delta(\mathbf{r}-\mathbf{r}_i)\sigma_i v^{\mu}_i\equiv\rho w^{\mu},
\eea
are the currents for charge and number, respectively. 
 
We use  Eq.~\eqref{minimalcoupling} to  obtain the relation between $u$ and $v$ on  a  curved surface from it on a plane~\cite{Yu:2017siv} : 
\bea
\label{relation1}
\rho v^{\mu}=\rho u^{\mu}-\eta\epsilon^{\mu}_{\nu}g^{\nu \alpha}\nabla_\alpha \sigma, \\
\rho w^{\mu} =\sigma u^{\mu}-\eta \epsilon^{\mu}_{\nu}g^{\nu \alpha}\nabla_\alpha \rho.
\label{relation2}
\eea
Consequently,  $\omega_v-\omega =\eta \nabla_\mu (\frac{1}{\rho}\nabla^\mu\sigma)$, where 
$\omega_v=\epsilon^\gamma_\alpha \nabla_\gamma v^\alpha$ and  $\omega=\epsilon^\mu_\nu \nabla_\mu u^\nu=8\pi \eta \sigma $.  The vortex fluid is compressible and  $\nabla_\mu v^\mu =-\eta \epsilon^\mu_\nu\nabla_\mu \left(\frac{1}{\rho}\nabla^\nu \sigma\right) \neq 0$ \cite{footnoteint}.  Note that here $\epsilon^\beta_\alpha \equiv
\epsilon^\beta_{\,\,\,\,\alpha} =g^{\beta \gamma} \epsilon_{\gamma \alpha}$ and $\epsilon_{\gamma \alpha} =\sqrt{\det g_{\mu \nu} }\tilde {\epsilon}_{\gamma \alpha}$ is the Levi-Civita tensor and $\tilde {\epsilon}_{\gamma \alpha}$ is the Levi-Civita symbol. In isothermal co- ordinates, the tensor $\epsilon^\beta_\alpha$ used here takes the same value as what is introduced previously [below Eq.~\eqref{tau}]. 
For a scalar $f$,  $\nabla_{\alpha}f =\partial_\alpha f$,  in complex coordinates, Eqs.~\eqref{relation1}~\eqref{relation2} become 
\bea
\label{coreformula1}
\rho v=\rho u-2i\eta\frac{1}{h}\partial_z\sigma, \\
\label{coreformula2}
\rho w =\sigma u-2\eta i\frac{1}{h}\partial_z\rho.
\eea
The above relations reveal that the velocity of a vortex at position $\mathbf{r}$ is the fluid velocity excluding the flow generated by the vortex itself at $\mathbf{r}$.  The superfluid velocity field $u$ is irregular at a vortex core and subtracting the pole at the vortex core  leads to a regular vortex velocity field $v$.

There is no solid reason  why the MC principle must lead to correct results~\cite{reall2013part}. Hence  justification is needed.  
To verify Eqs.~\eqref{coreformula1}\eqref{coreformula2}, let us apply the relation \eqref{coreformula1}, which is for  coarse-grained variables,  to discrete point vortices. 
The fluid velocity generated by these point vortices on a closed surface  is $u=2i h^{-1} \partial_z \psi$ with the stream function 
\bea \psi(z)=8\pi  \eta \sum_{i} \sigma_i G(z,z_i),
\eea
 where $G(z,z_i)$ is the Green's function satisfying~\cite{dritschel2015motion}
 \bea 
 \Delta G(z,z_i)=-\delta_{z,z_i}+1/\Omega,
 \eea
$\Delta \equiv \nabla^{\mu} \nabla_{\mu}$,  $\Omega$ is the area of the surface and 
 \bea
 \delta_{z,z_k}\equiv h^{-1}\delta(z-z_k).
 \eea
 The fluid velocity at $z=z_k$ is 
\bea
\hspace{-5.0mm}u_{z\rightarrow z_k}= \frac{16\pi  \eta  i}{h} \left[\sum_{i\neq k} \sigma_i \partial_z G(z,z_i) |_{z=z_k} +\sigma_k\lim_{z\rightarrow z_k}\partial_z G(z,z_k)  \right],
\label{singular}
\eea
where the last part is  the contribution from the vortex at $z=z_k$ itself  and contains a pole.  To analyze the last term in Eq.~\eqref{singular},  it is useful to isolate the logarithmic singularity of the Green's function~\cite{gustafsson2022vortex}:  
\bea
G(z,z_k)=\frac{1}{2\pi}\left[-\log|z-z_k|+H(z,z_k)\right],
\eea 
 where $H(z,z_k)=H(z_k,z)$ is a regular function.   Expanding  in a power series in $z$ around $z_k$, we obtain
  \bea
  H(z,z_k)=h_0(z_k)+ \frac{1}{2}h_1(z_k) \, (z-z_k)+{\rm H.c.}+{\mathcal O}(|z-z_k|^2) \nn\\
  \eea
  and 
   \bea
   \partial_z H(z,z_k)=\frac{1}{2}h_1(z_k)+{\mathcal O}(|z-z_k|)=\frac{1}{2}\partial_{z_k} h_0(z_k)+{\mathcal O}(|z-z_k|). \nn\\
   \eea
Here $h_0(z_k)=H(z_k,z_k)$ and $h_1(z_k)=\partial_z H_{z,z_k}|_{z=z_k}$.  Let us now  analyze the singular term in $\partial_z \sigma $.  Noting that $2/(\pi h) \partial_{\bar{z}}  \partial_z \log|z-z_k|=\delta_{z,z_k}$ and re-arranging derivatives , we obtain  
\bea
\label{singular2}
\lim_{z\rightarrow z_k}\partial_z \sigma=-\sigma_k \delta_{z,z_k} \partial_z \log h|_{z=z_k}-2\sigma_k  \frac{1}{z-z_k} \delta_{z,z_k},
\eea
where we have used  $\partial_{\bar{z}} (1/z)=\pi \delta(\mathbf{r})$.  
Hence the singular terms $\propto 1/(z-z_k)$ in  Eq.~\eqref{singular}  and  Eq.~\eqref{singular2} cancel and the remaining  finite part  in Eq.~\eqref{coreformula1}  gives rise precisely, by recognizing $v(z=z_k)=d \bar{z}_k(t)/dt$ and $\lim_{z\rightarrow z_k} \rho=\delta_{z,z_k}$, the EOM of point vortices on closed surfaces with zero genus~\cite{dritschel2015motion}:
\bea
\sigma_k h \frac{d \bar{z}_k(t)}{dt}=8 \pi \eta  i \left[2 \sum_{i\neq k} \sigma_k \sigma_i \partial_z G(z,z_i) |_{z=z_k} +  \partial_{z_k} R_{\rm robin}(z_k)\right], \nn\\
\label{PVMsurface}
\eea
where  $R_{\rm robin}(z_k) \equiv (1/2\pi) [h_0 (z_k)+\log \sqrt{h(z_k)}]$ is the celebrated Robin function~\cite{gustafsson2022vortex}.

Note that Eq.~\eqref{PVMsurface} holds for infinitely large curved surfaces as well~\cite{Hally}, and hence  so do Eqs.~\eqref{coreformula1}\eqref{coreformula2}. For an infinitely large surface, $R_{\rm robin}(z_k)= (1/2\pi) \log \sqrt{h(z_k)}$.
In contrast to the scenario  on a plane,  on a curved  surface  the self-energy of a vortex  is position dependent and  a single vortex may move driven by the geometrical potential (Robin function)~\cite{Turner2010}.  It was not  an easy task to obtain the EOM of point vortices on closed surfaces~\cite{dritschel2015motion}.  From the vortex fluid point of view,  it is somewhat striking that  relation~\eqref{coreformula1} naturally generalized from it on a plane  could  lead to Eq.~\eqref{PVMsurface}. 

\section{Dynamical equations of vortex fluids on closed  surfaces}
The Euler equation on a  curved surface  can be obtained from its form on a plane applying the  MC principle~\cite{Reuther2015, Mickelin18}:  
\bea
\partial_tu^\alpha+ \nabla_{\beta} {\cal T}^{\alpha \beta} =0,
\label{Euler}
\eea 
where  the momentum flux tensor $ {\cal T}^{\alpha \beta}= u^{\alpha}u^{\beta}+p g^{\alpha \beta}$ (hereafter we set the  fluid (mass) density $n=1$).   Unlike the case of Euler equation,  we can not  apply the MC principle  to Eq.~\eqref{vortexfluid} directly.   
The reason is that there are  terms containing  second derivatives of vectors in  Eq.~\eqref{vortexfluid}.  On a plane,  the order of  derivatives of  these terms are interchangeable, namely:  $\partial_\beta \partial^\alpha \partial^\beta \sigma=\partial^\alpha  \partial_\beta \partial^\beta \sigma$ and  $\partial_\beta \partial^\gamma v^{\alpha}= \partial^\gamma\partial_\beta v^{\alpha}$.  However  on a curved surface,  $\nabla_\beta \nabla^\alpha \nabla^\beta \sigma \neq \nabla^\alpha  \nabla_\beta \nabla^\beta \sigma$, and $\nabla_\beta \nabla^\gamma v^{\alpha}\neq \nabla^\gamma \nabla_\beta v^{\alpha}$.  At this stage, there is no preferred order for which  the MC substitution should be applied. 

Our strategy is to  search for another tensor  $Q^{\alpha \beta}$ such that 
\bea
&& 1) \text{ it does not contain derivatives of vectors};  \nn\\
&& 2) \, \partial_\beta T^{\alpha\beta}=\partial_{\beta} Q^{\alpha\beta}.\nn
\label{condition}
\eea
To do so, it is convenient to use  complex coordinates, in which, Eq.~\eqref{vortexfluid} becomes 
\bea
\partial_t (\rho v)+\partial_z T_{z\bar{z}}+\partial_{\bar z} T +\rho \partial_z (2p)=0,
\eea
where 
\bea
T&&=\rho vv+4\eta^2\sigma\partial_z\left(\frac{1}{\rho}\partial_z\sigma\right)-4i\eta\sigma\partial_z v ,\\
T_{z\bar{z}}&&=\rho v\bar{v}+16\pi\eta^2\sigma^2+4\eta^2\sigma \partial_{\bar{z}}\left(\frac{1}{\rho}\partial\sigma\right) \nn\\
&&= \rho v\bar{v}+4i\eta\sigma\partial_{\bar{z}}v-4\eta^2\sigma \partial_{\bar{z}}\left(\frac{1}{\rho}\partial\sigma\right). 
\eea
In the last step  we have used $\partial_{\bar z} u=-4 \pi i \eta \sigma$ and $u=v+2i \eta \partial_z\sigma/\rho$ ~\cite{Yu:2017siv}.  

Let us now define 
 \bea
  \label{Qtensor1}
 Q_{z\bar{z}}&\equiv\rho v\bar{v}-4i\eta  v\partial_{\bar{z}}\sigma+4\eta^2\frac{1}{\rho}\partial_{\bar{z}}\sigma\partial\sigma,\\
 Q&\equiv \rho vv+4i\eta v\partial_z\sigma-4\eta^2\frac{1}{\rho}\partial_{z}\sigma\partial_{z}\sigma.
 \label{Qtensor2}
 \eea
 Clearly condition 1)  is satisfied.  Since  $T_{z\bar{z}}-Q_{z\bar{z}}=4i\eta \partial_{\bar{z}}(\sigma v)-4\eta^2\partial_{\bar{z}}[(\sigma/\rho)\partial_z\sigma]$ and 
$T-Q=-4i\eta\partial_z(\sigma v)+4\eta^2\partial_z[(\sigma/\rho)\partial_z\sigma]$, it is easy to verify that 
$ \partial_z Q_{z\bar{z}}+\partial_{\bar{z}} Q=\partial_z T_{z\bar{z}}+\partial_{\bar{z}} T$
 which is the complex form of condition 2). Hence $Q^{\alpha \beta}$  defined in Eqs.~\eqref{Qtensor1} \eqref{Qtensor2} is  the tensor we search for.
 
 It is now ready to apply the MC principle to obtain the vortex fluid equation on a closed  surface :
  \bea
  \partial_t(\rho v^\alpha)+\nabla_\beta Q^{\alpha\beta}+\rho\nabla^\alpha p=0
  \label{AnomalousCartesian}
  \eea
 where 
  \bea 
  Q^{\alpha \beta}=\rho v^\alpha v^\beta+2\eta  v^\alpha \epsilon^\beta_\mu\nabla^\mu \sigma +\eta^2 \frac{1}{\rho} \epsilon^{\alpha}_\mu \epsilon^{\beta}_\nu \nabla^{\mu} \sigma \nabla^\nu \sigma
  \eea
 and the pressure $p$ is determined by $\nabla_{\mu }(u^{\nu}\nabla_{\nu}u^{\mu})=-\nabla_{\mu}\nabla^{\mu}p$. 
 
It is crucial  that the momentum flux tensor includes the odd viscous tensor $\tau^{\alpha\beta}(\partial_{\mu} \rightarrow \nabla_{\mu})$. For this purpose, we need to write the dynamical equation in terms of $T^{\alpha \beta}$:
\bea
\partial_t \left(\rho v^\alpha\right)+\nabla_\beta T^{\alpha \beta}+\rho\nabla^\alpha p=\eta K\left(\eta \frac{\sigma}{\rho}\nabla^\alpha\sigma-2\sigma \epsilon^\alpha_\beta v^\beta \right),
\label{vortexfluideq}
\eea
where 
\bea
T^{\alpha \beta}=\rho v^\alpha v^\beta+\eta^2\sigma\nabla^\beta(\frac{1}{\rho}\nabla^\alpha\sigma)+8\pi\eta^2\sigma^2g^{\alpha\beta}+\sigma\tau^{\alpha\beta}, 
\eea
$K=R_{1212}/\det g_{\mu \nu}=R_{1212}/h^2$ is  Gaussian curvature. 
Here we have used $\epsilon^{\mu}_{\nu} \nabla _{\mu} u^{\nu}=8\pi \eta \sigma$ and Eq.~\eqref{relation1}.  

Comparing to Eq.~\eqref{vortexfluid},  the conspicuous feature  of  Eq.~\eqref{vortexfluideq} is that  the combination of Gaussian curvature and the circulation quantum/odd viscosity plays the role of the coefficient of a source term.  The presence of this additional term  might be referred to as {\em curvature anomaly}.   The momentum flux tensor  $T^{\alpha \beta}$ is not symmetric for binary vortex fluids and it can not be symmetrized  in the usual way due to that its anti-symmetric part  
\bea
T^{1 2}-T^{21}=\eta^2\sigma h^{-1} \nabla_\mu v^\mu
\eea
 is not a total divergence.  
The hydrodynamics equation \eqref{vortexfluideq}  is invariant under the following scaling transformation  
$x\rightarrow \lambda x$, $t \rightarrow \lambda^2 t$, $\rho \rightarrow \lambda^{-2} \rho$ , $\sigma \rightarrow \lambda^{-2} \sigma$, $v^{\alpha} \rightarrow \lambda^{-1} v^{\alpha}$ $K\rightarrow \lambda^{-2} K$, $p\rightarrow \lambda^{-2} p$.
The vortex core size $\ell$ plays the role of the ultraviolet cut-off of the hydrodynamics theory.

Since the odd  viscous tensor $\tau^{\alpha \beta}$ is of fundamental impotence  and appears in a large class of fluids~\cite{oddviscosity} , it is worthwhile exploring  its properties on a curved surface.  From the definition of $\tau^{\alpha \beta}$, one obtains 
\bea
\Delta v_\alpha \nabla_{\beta} \tau^{\alpha \beta}=-\eta K   \epsilon^{\alpha}_\beta v^\beta \Delta v_\alpha.
\eea 
For a closed orientable surface, due to Gauss-Bonnet theorem, we have  
\bea
\int ds \frac{\Delta v_\alpha \nabla_{\beta} \tau^{\alpha \beta}}{\epsilon^{\alpha}_\beta v^\beta \Delta v_\alpha }=-\eta \int ds K =-2\pi \eta \chi ({\cal M}),
\label{topology}
\eea
where $\chi ({\cal M})=2(2-g)$ is Euler characteristic, and $g$ is the genus of the surface.   It should be noted that Eq.~\eqref{topology} holds for any value of $g$.  Connecting Eq.~\eqref{topology} to physical observable  deserves future investigations. 
The hydrodynamic equation~\eqref{vortexfluideq}  can be verified by substituting  Eqs.~\eqref{relation1}~\eqref{relation2} into Eq.~\eqref{Euler}.

\section{Vortex flow on  a sphere} We consider vortex fluids on a sphere embedded in $\mathbb{R}^3$.  We introduce the Cartesian coordinates 
\bea
\xi=R \sin \theta \cos \phi, \quad  \eta=R \sin \theta \sin \phi, \quad \zeta=R \cos \theta,
\eea
  where $R$ is the radius,  $\theta$ is the polar angle and  $\phi$ is the azimuthal angle.   On a sphere, stereographic coordinates $z=x^1+i x^2$ are isothermal coordinates and  are  related to the spherical coordinates by
$z=R\tan (\theta/2) e^{i\phi}$ (projection from the south pole). In terms of $z$, the Riemannian metric reads
\bea
h=\frac{4 R^4}{(R^2+|z|^2)^2}
\eea
 and in spherical coordinates  
\bea
ds^2=R^2d\theta^2+R^2\sin^2\theta d\phi^2.
\eea
\subsection{Conserved quantities}
It is known that  for point vortices  on a sphere,  the  quantities 
\bea
L_{\xi}&&=\kappa R^2\sum_j \sigma_j \sin\theta_j \cos\phi_j, \\
L_{\eta}&&=\kappa R^2 \sum_j \sigma_j \sin\theta_j \sin\phi_j, \\
L_{\zeta}&&=\kappa R^2 \sum_j \sigma_j \cos\theta_j,
\eea
are conserved~\cite{Kimura99}. 
In terms of collective variables, 
\bea
L_{\xi}&&=\kappa R^2 \int ds\, \sigma \sin\theta \cos\phi , \\
 L_{\eta}&&=\kappa R^2\int ds\, \sigma \sin\theta \sin\phi,  \\
 L_{\zeta}&&=\kappa R^2 \int  ds\, \sigma \cos\theta.
\eea 
These conserved quantities are directly related to the corresponding fluid angular momentum  $\int ds \, \mathbf{r} \times \mathbf{u}$  which  is associated with the $\textrm{SO}(3)$ symmetry.  In  stereographic coordinates, they become  
\bea
L_{\xi}&&=\kappa R\int  dx^1 dx^2 h^{3/2}\sigma  x^1, \quad L_{\eta}=\kappa R\int dx^1dx^2 h^{3/2} \sigma  x^2, \\
L_{\zeta}&&=-\frac{\kappa}{2} \int  dx^1 dx^2 h^{3/2}\sigma |z|^2
+\frac{\kappa R^2}{2} \int  dx^1 dx^2 h^{3/2} \sigma.
\eea  
Then it is easy to notice that, as $R\rightarrow \infty $,  

\bea 
L_{\xi}/R  \propto  P_{x^2}&&=-\kappa \sum_i\sigma_i x^1_i=-\kappa \int dx^1 dx^2 \sigma x^1, \\
L_{\eta}/R \propto P_{x^1}&&=\kappa \sum_i\sigma_i x^2_i=\kappa \int dx^1 dx^2\sigma x^2, 
\eea
where  $P_{x^2}$ and  $P_{x^1}$ are components of  canonical momentum of  vortices on a plane. 
Also,  as $R\rightarrow \infty $, 
\bea 
L_{\zeta} \propto L =\kappa \sum_i\sigma_i|\mathbf{r}_i|^2=\kappa \int dx^1 dx^2 \sigma |z|^2
\eea
 which is the canonical angular momentum of the point-vortex system on a plane.  Hence there is a one-to-one correspondence between conserved quantities on a plane and on a sphere.

It is worthwhile  to mention that the enstrophy  
\bea H \equiv \int ds \, \omega^2
\eea 
 is conserved in  any closed surface with zero genus, as 
\bea 
\frac{dH}{dt}=-2 \int ds \omega    u^\mu\nabla_\mu \omega =- \int ds \, \nabla_\mu (  u^\mu  \omega^2) =0.
\eea
However  the symmetry associated with this conservation law is not obvious~\cite{Enstrophysymmetry}.

\subsection{Stationary vortex flows}
For constant vortex density $\rho=\rho_0$ on a surface with constant Gaussian curvature $K=K_0$,  the vortex fluid becomes  incompressible $\nabla_\mu v^\mu=0$ and Eq.\eqref{vortexfluideq}  becomes 
\bea
\label{vortexfluideq2}
\partial_t \omega_v+\frac{1}{\rho_0}\epsilon^\gamma_\alpha \nabla_\gamma \nabla_\beta T^{\alpha \beta}=\frac{2\eta K_0}{\rho_0}  v^\beta \nabla_\beta \sigma.   
\eea

For a sphere, $K_0=1/R^2$, and we find a stationary solution of Eq.~\eqref{vortexfluideq2}
\bea
\sigma&&=\rho_0\frac{K^{-1}_0-\abs{z}^2}{K^{-1}_0+\abs{z}^2}, \\
\quad v^1&&=-\left(4\pi \eta\rho_0-K_0\eta\right)x^2, \quad v^2=\left(4\pi \eta\rho_0-K_0\eta\right)x^1. 
\eea
Note that $\sigma(z=0)=\rho_0=-\sigma(z=\infty)$. 
For this flow $\tau^{\alpha \beta}=0$, $L_{\xi}=L_\eta=0$ and $L_{\zeta}=4/3 \pi R^4 \kappa \rho_0$.  The modulus of the vortex velocity field is 
\bea
|v|=\sqrt{v_1 v^1+v_2 v^2}= \frac{2 R^2 |4\pi \eta \rho_0-K_0\eta| |z|}{R^2+|z|^2},
\eea
having the profile of a Kaufmann vortex. For  $|z| \ll R$, $|v| \propto  |z|$, while $|v| \propto  1/|z|$  for $|z| \gg R$.  The maximum value of $|v|$ is reached at $|z|=R$.  The  anomalous correction  to the fluid velocity is 
\bea
\label{correctionvelocity}
v^1-u^1=K_0 \eta x^2, \quad v^2-u^2=-K_0 \eta x^1   
\eea
and its modulus is  
$(v^1-u^1)(v_1-u_1)+(v^2-u^2)(v_2-u_2)=h K^2_0 \eta^2 |z|^2=4K_0 \eta |z|^2/(R^2+|z|^2)^2$.
The vorticity of the vortex velocity field also has an anomalous correction that is proportional to $K_0$ 
\bea
\label{correctionvorticity}
\omega_v-\omega=-2K_0\eta \frac{K^{-1}_0- |z|^2}{K^{-1}_0+ |z|^2}.
\eea
 When $R\rightarrow \infty$, $K_0\rightarrow 0$,  $\sigma \rightarrow \rho_0$ for  $z \neq \infty$ ($\rho_0$ keeps a constant as $R \rightarrow \infty$), and this corresponds to  rigid body rotation of a chiral vortex flow on a plane.  The oppositely charged vortices accumulate at $z=\infty$.  It is important to note that the anomalous  corrections, i.e., the differences between $v$ and $u$ (or $\omega_v$ and $\omega$), are proportional to curvature and  vanish as $K_0 \rightarrow 0$.

\begin{figure}[htp] 
	\centering
	\includegraphics[width=0.45\textwidth]{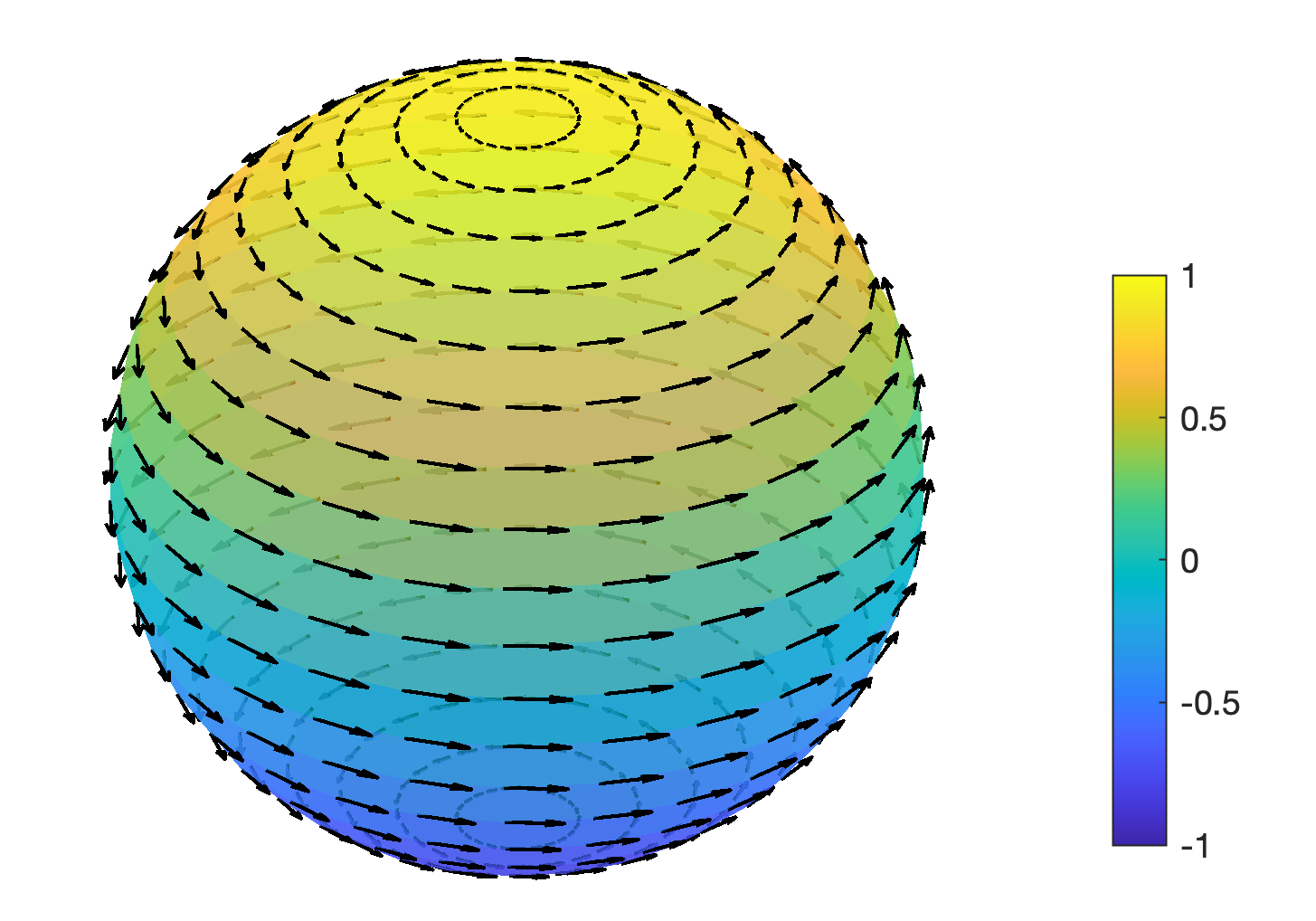}
	\caption{Schematic of the stationary vortex flow on a sphere.  The arrows represent  the vortex velocity field $\mathbf{v}$ and the background color shows the renormalized vorticity of the vortex fluid $\omega_v(\theta)/|\omega_v(0)|$. 
	\label{f:profile}}
\end{figure} 

It is helpful to express this stationary flow using spherical coordinates, for which 
$\mathbf{v}=v^{\theta} \partial_{\theta}+v^{\phi} \partial_{\phi}$ and 
\bea
\sigma=\rho_0 \cos\theta,\quad v^\phi=R(4 \pi \eta\rho_0-K_0 \eta)  , \quad v^\theta=0.
\eea
The modulus of the vortex velocity field is
\bea
|\mathbf {v}|=\sqrt{ v^{\phi}v_{\phi}+ v^{\theta}v_{\theta}}=R|(4\pi \eta \rho_0-K_0\eta) \sin  \theta \,|,
\eea
which  vanishes at the poles and reaches the maximum at the equator (see Fig.~\ref{f:profile}). 
Since $u^\phi=4\pi \eta R \rho_0 $ and $u^\theta=0$,  we have  $|v-u|^2=(v^{\phi}-u^{\phi})(v_{\phi}-u_{\phi})=K_0 \eta^2 \sin^2 \theta$. 
The vorticity  of the vortex fluid reads 
\bea
\omega_v
&&=2\left(4\pi\rho_0-K_0\right)\eta \cos\theta.
\eea
and the  correction is $\omega_v-\omega=-2K_0 \eta \cos \theta$.  Due to  compactness of the sphere,  $\omega_v(\theta=0)=-\omega_v(\theta=\pi)=2\left(4\pi\rho_0-K_0\right)$,  the vorticity of this vortex flow has the profile of a vortex-dipole. 
It is worthwhile mentioning that the vortex flows we found here are analogous to zonal Rossby–Haurwitz  flows in Euler fluids on a sphere~\cite{tritton2012physical, nualart2023zonal},  which play an important role in analyzing dynamics of  atmosphere of Earth~\cite{rossby1939relation, platzman1968rossby, haurwitz1940motion}.

\section{Conclusion}
We generalize the vortex fluid theory on a plane to closed surfaces  with zero genus.  The dynamical equation is derived  using the minimal coupling principle  from it on a flat surface.  An additional  curvature term emerges and describes the interaction between topological defects and curvature in the hydrodynamical level. 
Since the vortex fluid equation contains second derivatives of vectors,  there is an ambiguity for  applying the minimal coupling principle directly. 
Our method does get over this difficulty  and  provides a feasible recipe to investigate other complex fluids on curved surfaces. 
For a sphere,  a nontrivial stationary vortex flow is found analytically.  It poses a challenge to find analytic solutions for surfaces with non-constant Gaussian curvature. For surfaces with Gaussian curvature weakly depending on positions [$K(x^\mu)=K_0+\delta K(x^{\mu})$ with $\delta K(x^{\mu}) \ll K_0$], it might be  possible to treat $\delta K(x^{\mu}) $ as perturbations and investigate the effects of non-constant Gaussian curvature. These are interesting topics which are worthwhile exploring in the feature. 

The theory developed in this work leads to a broad  understanding of the interaction between topological defects and curvature,  and provides a theoretical  framework  for investigating  rich  phenomena involving a large number of quantum vortices~\cite{PhysRevResearch.4.013122, Kanai2021, GuangyaoPRA, Saito2023} 
in bubble trapped  Bose-Einstein condensates~\cite{Tononi2020,PhysRevLett.123.160403}.

\section*{Acknowledgment}
We acknowledge J. Frauendiener, B. Feng, A. M. Mateo,  L. A. Williamson, P. B. Blakie and A. S. Bradley for useful discussions. 
X.Y. acknowledges support from the National Natural Science Foundation of China (Grant No. 12175215), the National Key Research and Development Program of China (Grant No. 2022YFA 1405300) and  NSAF (Grant No. U2330401).


%

\end{document}